\documentclass[amsmath,amssymb,floatfix,preprint,showpacs,superscriptaddress]{revtex4-1}

\usepackage{graphicx}
\usepackage{dcolumn}
\usepackage{bm}
\usepackage{amsmath}
\usepackage{amsthm}
\usepackage{amsfonts}
\usepackage{epstopdf}
\usepackage{latexsym}\usepackage{array}
\usepackage{subfig} 
\usepackage{color}
\usepackage{float}

\begin{document}


\title{Ergodic transition in a simple model of \\ the continuous double auction}

\author{Tijana Radivojevi\'{c}}
\email{tradivojevic@bcamath.org}
\homepage{www.bcamath.org/en/people/radivojevic}
\affiliation{BCAM - Basque Center for Applied Mathematics, Alameda de Mazarredo 14, 48009 Bilbao, Basque Country, Spain}

\author{Jonatha Anselmi}
\email{anselmi@bcamath.org}
\homepage{www.bcamath.org/en/people/anselmi}
\affiliation{BCAM - Basque Center for Applied Mathematics, Alameda de Mazarredo 14, 48009 Bilbao, Basque Country, Spain}

\author{Enrico Scalas}
\email{enrico.scalas@unipmn.it}
\homepage{www.mfn.unipmn.it/~scalas}
\affiliation{Dipartimento di Scienze e Innovazione Tecnologica, Universit\`a del
Piemonte Orientale ``Amedeo Avogadro'', Viale Michel 11, 15121 Alessandria,
Italy and \\ BCAM - Basque Center for Applied Mathematics, Alameda de Mazarredo 14, 48009 Bilbao, Basque Country, Spain}

\date{\today}

\pacs{
05.40.Ca, 
05.45.Tp, 
05.10.Gg  
}


\begin{abstract}
\textcolor{black}{We study a phenomenological model for the continuous double auction, equivalent to two independent $M/M/1$ queues. The continuous double auction defines a continuous-time random walk for trade prices. The conditions for ergodicity of the auction are derived and, as a consequence, three possible regimes in the behavior of prices and logarithmic returns are observed. In the ergodic regime, prices are unstable and one can observe an intermittent behavior in the logarithmic returns. On the contrary, non-ergodicity triggers stability of prices, even if two
different regimes can be seen. 
}
\end{abstract}

\maketitle

{\color{black}{

\section*{Introduction}

The {\it continuous double auction} is the trading system used by regulated equity markets.
The auction is called {\it double} as demand and offer are collected in a book where orders to buy and to sell are registered, and {\it continuous} because orders can be placed at any instant in a given daily time window.
The detailed rules for trading may be different from one stock exchange to the other, but, essentially, things work as follows. 
Traders can either place buy orders ({\em bids}) or sell orders ({\em asks}) which are then registered in a {\em book} for a particular stock traded in the exchange. The {\em limit order} is the typical one.
A {\it bid limit order} is specified by two numbers: the quantity $q_b^{(T)}$ that trader $T$ wants to buy and the upper limit price $p_b^{(T)}$ she is willing to pay for a single share. An {\it ask limit order} is an order to sell $q_a^{(T)}$ units of the share at a price not smaller than a limit price $p_a^{(T)}$ selected by trader $T$. The couples $(p_{b}^{(T)},q_{b}^{(T)})$ and $(p_{a}^{(T)},q_{a}^{(T)})$  are written in the book and ordered from the best bid to the worst
bid and from the best ask to the worst ask, respectively. The {\em best bid} is the price $p_b = \mathrm{max}_{T \in I_b} (p_b^{(T)})$, where $I_b$ is the set of
traders placing bids, whereas the {\em best ask} is the price $p_a = \mathrm{min}_{T \in I_a} (p_a^{(T)})$, where $I_a$ is the set of traders placing asks. At every time $t$, one has that $p_a (t) > p_b (t)$. 
Occasionally, a {\em market order} may take place, when a trader accepts a best bid or best ask price from the book, and the $i$-th trade occurs at the epoch $t_i$. Stock exchanges specify rules for the priorities of limit orders placed at the same price and for execution of market orders with quantities that are not totally available at the actual best price. The sequence of prices $p(t_i)$ at which trades take place at epochs $t_i$ is an important process for understanding market dynamics. As detailed below, one can describe this sequence in terms of a suitable continuous-time random walk.}}

\section*{Model}

In our model, following \cite{cont10}, prices assume $N$ integer values from $1$ to $N$. A price can be regarded as a {\em class} where orders are placed. This way of representing prices is a faithful representation of what happens in real markets due to price discretization. The only unrealistic feature is
the presence of an upper limit to prices which we keep to ensure partial analytical tractability. Note that orders can be considered as {\em objects}
to be classified by prices at which they are placed (see \cite{garibaldi2010} for a general discussion on the problem of allocating objects to 
classes). We only consider two kinds of orders, namely, limit orders and market orders.
\begin{figure}[h!]
\includegraphics[width=0.7\columnwidth ]{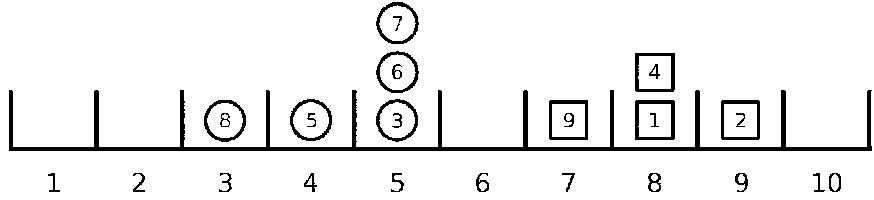} 
\label{book}
\caption{A state of the system. Limit bid orders are depicted by circles whereas squares represent limit ask. Each order can be described by a label. 
An individual description is a list showing, for each limit order, whether it is a bid or an ask and to which category (price) 
it belongs. Here, it is $x_1=(8,a), \ x_2=(9,a), \ x_3=(5,b), \ x_4=(8,a), \ x_5=(4,b), \ x_6=(5,b), \ x_7=(5,b), \ x_8=(3,b), \ x_9=(7,a)$.
A statistical description is a list giving us for each category (price) the number of limit orders either of bid or of ask type 
it contains. In this case
it is $y_1^a=\ldots=y_6^a=0, \ y_7^a=1, \ y_8^a=2, \ y_9^a=1, \ y_{10}^a=0$ for ask orders and $y_1^b=y_{2}^b=0, \ y_{3}^b=1, \ y_{4}^b=1, \ y_{5}^b=3, \ y_{6}^b=\ldots=y_{10}^b=0$ for bid orders. 
Finally, a partition description is the number of categories (prices) with zero limit ask or limit bid orders, one ask/bid limit order, etc.. In this case we have $z_0^a=7, \ z_1^a=2, \ z_2^a=1$ for ask orders and $z_0^b=7, \ z_1^b=2, \ z_2^b=0, \ z_3^b=1$
 for bid orders.}
\end{figure}
As discussed above, limit orders can be of two types: limit bid orders, i.e.\ orders to buy {\em one} share and  limit ask orders, i.e.\ orders to sell {\em one} 
share. Market orders
are also of two types as either the best limit bid order or the best limit ask order are accepted when a transaction occurs. When a market order arrives, only one limit order will be removed either from the best bid category or the best ask category, provided that these categories contain at least one order. In other words, we assume that all the orders are characterized by quantities equal to one share. We shall further assume
that limit ask orders arrive following a Poisson process with rate $\lambda_a$ and that limit bid orders arrive at a rate $\lambda_b$. These orders are
written in the book and they wait until a market order arrives. Remember that the best limit bid orders, namely the limit bid orders offered at the highest
price are always strictly smaller than the best limit ask orders, that is the limit ask orders offered at the lowest price. Market orders to buy, 
accepting one of the best ask orders, arrive in the market separated by exponentially distributed waiting times with parameter $\mu_b$, whereas
market orders to sell arrive with an exponential distribution of waiting times with parameter $\mu_a$. If the book is empty, or if the appropriate side of the book is empty, market orders are not 
executed. It is necessary to remark that order inter-arrival times are not exponentially distributed in real markets \cite{scalas06b} 
due to the non-stationary
behavior of humans \cite{barabasi10}. However, to ensure analytical tractability, here we assume that inter-arrival times are
exponentially distributed. We shall assume that limit ask orders are uniformly placed in the price classes from ${p}_b + 1$ to ${p}_b + n$, where
${p}_b$ is the class of the current best bids. Conversely, limit bid orders are uniformly placed in the price classes from
${p}_a - n$ to ${p}_a -1$, where ${p}_a$ is the class of the current best asks. As mentioned above, the accessible system states are
limited by the condition ${p}_b < {p}_a$. When ${p}_a$ is between $1$ and $n$ (${p}_b$ between $N-n+1$ and $N$), 
the bid (respectively, ask) interval is restricted correspondingly. For instance, if ${p}_a = 1$, no bids are possible.
Finally, if no orders are present in the book, the next bid, $b$,
will be uniformly chosen in $p-n \leq b \leq p$ and the next ask, $a$ in $p \leq a \leq p+n$, where $p$ is the price of the last trade.
The specification of an initial price (which can be interpreted as the opening auction price) is then sufficient to start the auction.
Note that the model outlined above is essentially the same as in \cite{smith02}. In terms of agent-based models, it is a
{\em zero intelligence agent-based model} \cite{gode}. However, our version does not suffer from using odd mathematical objects such as
uniform distributions over semi-infinite intervals. A preliminary discussion of our model was presented in \cite{RAS}.

The trade price process $P(t)$ is a continuous-time random walk that we wish to characterize. Let $T_i$ denote the epoch of the $i$-th trade; 
in particular, we are interested in the behaviour of the following random variable
\begin{equation}
\label{logreturn}
R_i = \log(P(T_{i+1})/P(T_i)),
\end{equation}
called the {\em tick-by-tick} logarithmic return. This is the usual variable used in statistical finance and financial econometrics for the
analysis of tick-by-tick data \cite{scalas06a}. It turns out that the behaviour of $R_i$ crucially depends on the presence or absence of statistical 
equilibrium in the supply mechanism.

\section*{Main results}

Indeed, there are two main regimes in this model and they are triggered by an ergodic transition.
Let us denote by $A(t)$ the total number of limit ask orders and by $B(t)$ the total number of limit bid 
orders present in the book. By definition, these two random processes are independent and they are $M/M/1$ queues with rates
$\lambda_a$ and $\mu_b$ and $\lambda_b$ and $\mu_a$, respectively \cite{queue}. $M/M/1$ queues are the
continuous-time equivalent of birth and death Markov chains. The conditions for the existence 
of statistical equilibrium (ergodicity) are given by the following inequalities
\begin{equation}
\label{askqueue}
\lambda_a < \mu_b, 
\end{equation}
and
\begin{equation}
\label{bidqueue}
\lambda_b < \mu_a.
\end{equation}
The intuitive meaning of these conditions is as follows. If the rate of arrival for limit orders is larger than the rate of market orders, then the number of orders in the book eventually explodes. However, in this case, prices will be able to fluctuate only among a few values. When the rate of market orders is larger than the rate of limit orders, the number of orders in the book remains finite and prices are free to 
fluctuate over the whole available range. In the ergodic regime, the invariant (and equilibrium) distributions of $A(t)$ and $B(t)$ are given by
two geometric distributions
\begin{equation}
\label{askdistribution}
\mathbb{P}(A = a) = \left( \frac{\lambda_a}{\mu_b} \right)^a \left(1 - \frac{\lambda_a}{\mu_b} \right),
\end{equation}
and
\begin{equation}
\label{biddistribution}
\mathbb{P}(B = b) = \left( \frac{\lambda_b}{\mu_a} \right)^b \left(1 - \frac{\lambda_b}{\mu_a} \right).
\end{equation}
Given the independence between $A(t)$ and $B(t)$, the joint probability density is
\begin{equation}
\mathbb{P} (A = a, B= b) = \left( \frac{\lambda_a}{\mu_b} \right)^a \left(1 - \frac{\lambda_a}{\mu_b} \right) \left( \frac{\lambda_b}{\mu_a}
\right)^b \left(1 - \frac{\lambda_b}{\mu_a} \right),
\end{equation}
from which one can find the probability of finding an empty book
\begin{equation}
\label{emptybook}
\mathbb{P} (A = 0, B= 0) = \left(1 - \frac{\lambda_a}{\mu_b} \right) \left(1 - \frac{\lambda_b}{\mu_a} \right).
\end{equation}

\begin{figure}[h]
\begin{center}
\subfloat[]{\includegraphics[width=0.33\columnwidth]{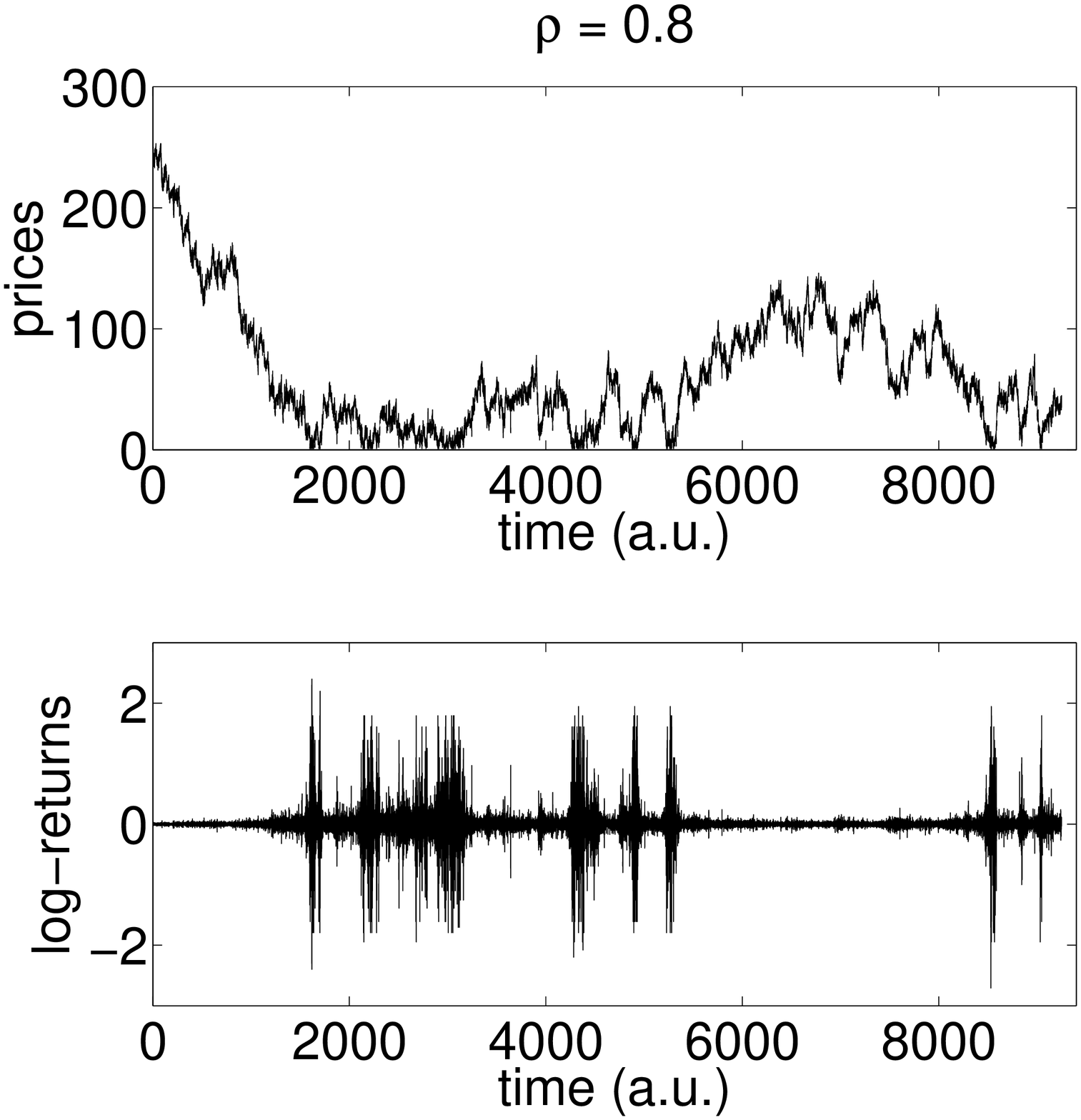} } 
\subfloat[]{\includegraphics[width=0.33\columnwidth]{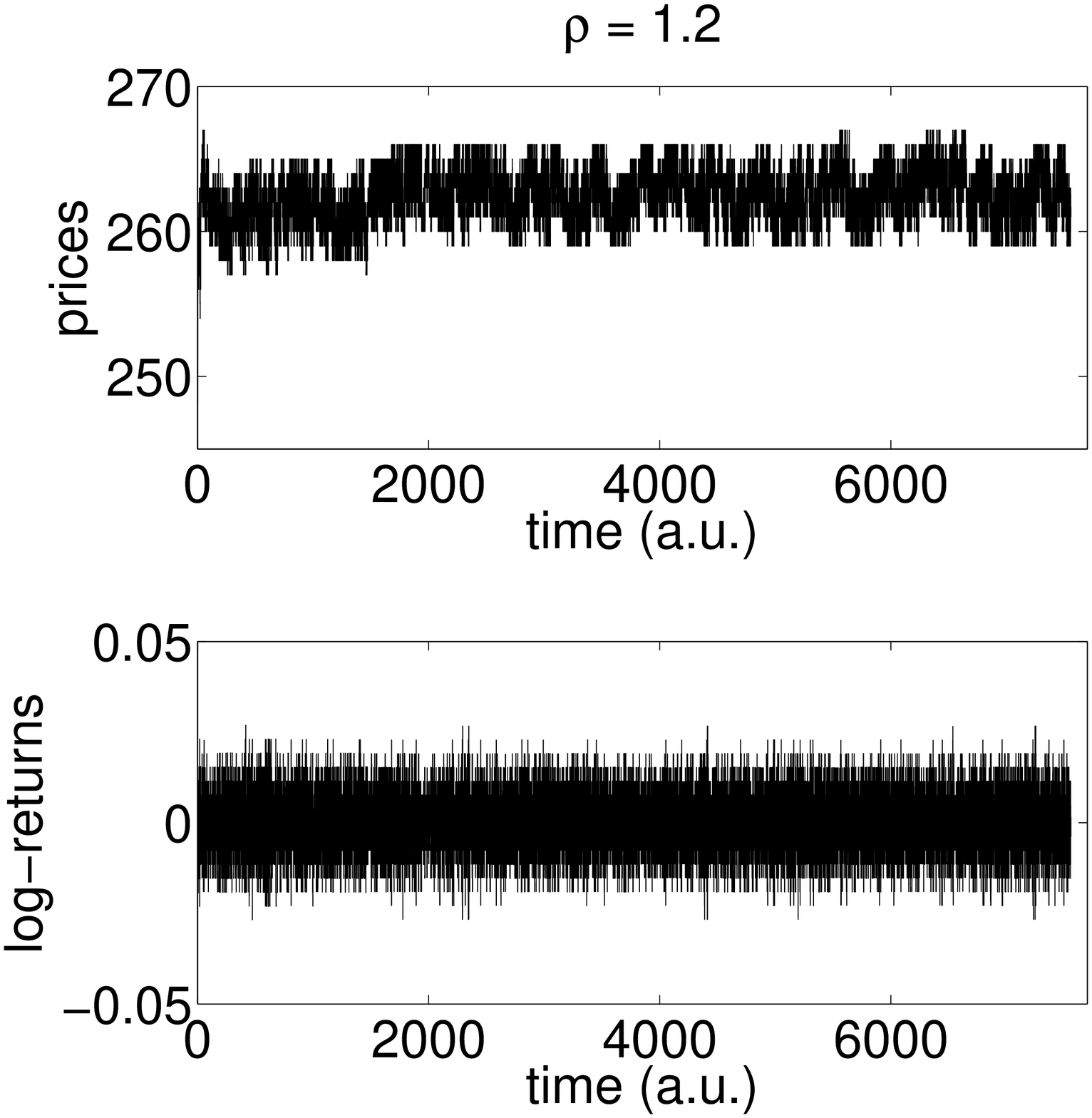} } 
\subfloat[]{\includegraphics[width=0.33\columnwidth]{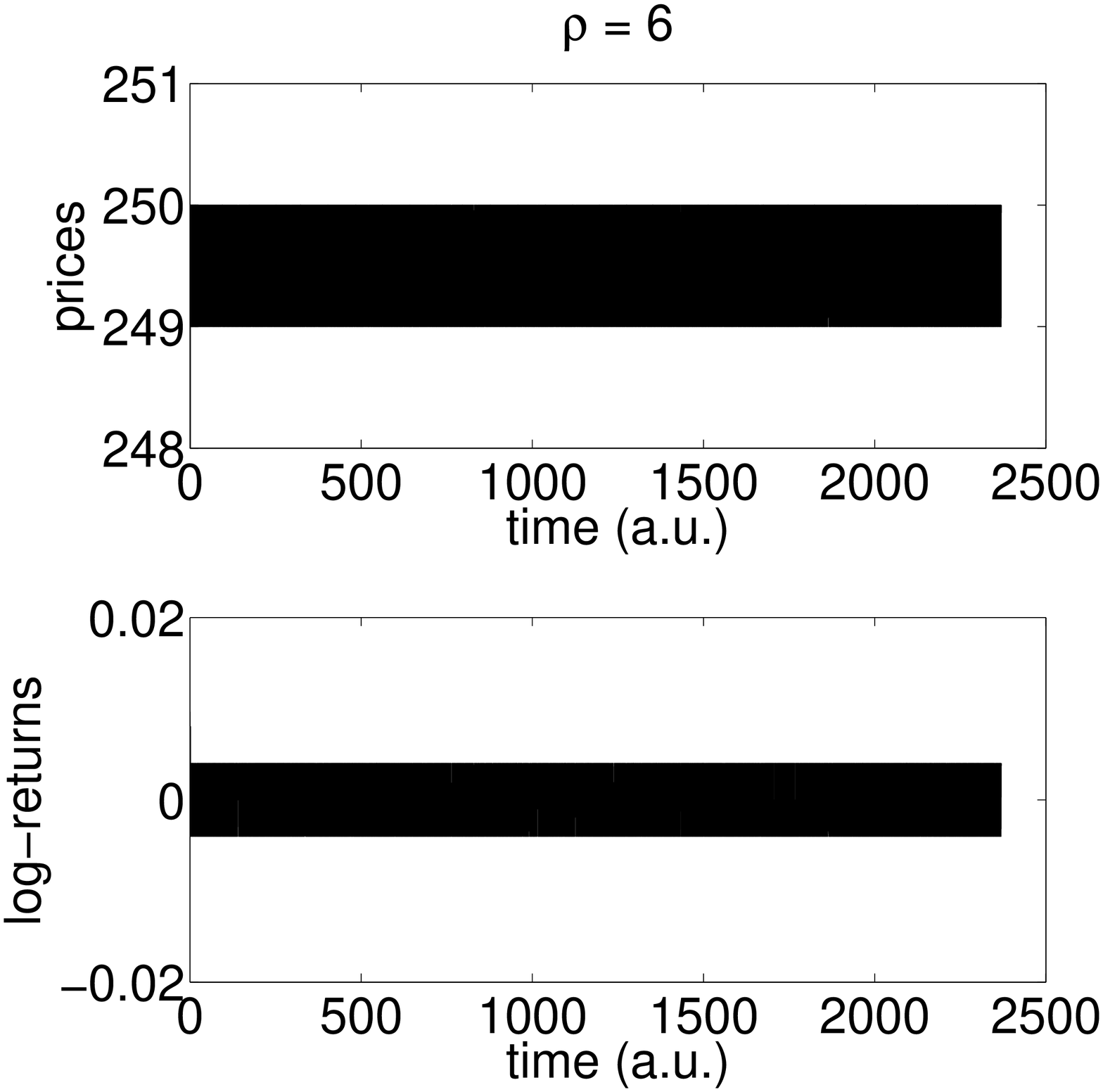} }
\caption{Time series of prices and log-returns in a system of $N=500$ prices, length of interval for placing orders $n=5$, with initial price $p_0=250$ and number of simulated events $t=10^5$. (a) ergodic case ($\rho=0.8$); (b) non-ergodic case for $1\leq \rho <n$ ($\rho=1.2$) and (c) non-ergodic case for $\rho \geq n$ ($\rho=6$).}
\label{timeseries}
\end{center}
\end{figure}

For our further analysis, we shall  focus on the case of a symmetric auction, assuming $\lambda_a = \lambda_b = \lambda$ and $\mu_a = \mu_b = \mu$ and we shall consider the ratio $\rho = \lambda / \mu$ as the basic order parameter of the model. In fact, there is no reason for a random
auction to be unbalanced towards selling or buying. 
As discussed above, if $\rho < 1$, we are in the ergodic regime, whereas for $\rho \geq 1$,
we are in a regime where the orders accumulate and $A(t), B(t) \to \infty$, 
for $t \to \infty$. The two regimes give rise to two radically different 
behaviours for the tick-by-tick log-returns \eqref{logreturn}. This is qualitatively shown in Fig. \ref{timeseries}, where we report the behavior of prices and log-returns in a Monte Carlo simulation for $\rho = 0.8$ (Fig. \ref{timeseries} a),  $\rho = 1.2$ (Fig. \ref{timeseries} b) and for $\rho = 6$ (Fig. \ref{timeseries} c). One can see by eye that, in the ergodic regime, high and low
log-returns are clustered, whereas, in the non-ergodic one, such a volatility clustering does not occur. Fig. \ref{timeseries} a clarifies the origin of clustering. When the
price is lower, log-returns are higher and the price process has the persistence behavior typical of random walks which immediately leads to clusters of low and high volatility as the price slowly moves up and down, respectively. The comparison between Fig. \ref{timeseries} b and Fig. \ref{timeseries} c
shows that there are two sub-regimes in the non-ergodic case. If $1 \leq \rho < n$, even if $A(t)$ and $B(t)$ diverge, the limit orders belonging to the best bid and the best ask can be removed by market orders and prices can fluctuate among a set, whereas if $\rho \geq n$, then after a transient, the number of limit
orders belonging to the best bid and the best ask diverges and prices can only fluctuate between two values. In this condition, the price process becomes a random telegraph process.
This behavior is justified by the fact that the process of the number of orders at the best bid price (respectively, best ask) can be coupled with the state of an $M/M/1$ queue with arrival rate
$\lambda_b/n$ ($\lambda_a/n$) and service rate $\mu_a$ ($\mu_b$); this is so because limit orders, upon arrival, distribute uniformly over the $n$ best prices.
If $\lambda_b/(\mu_a n) = \rho/n \ge 1$, then the number of orders at the best bid price converges to infinity, 
as $t \to \infty$, meaning that all trades will occur at the price where the bids accumulate.
If $\lambda_b/(\mu_a n)= \rho/n <1$, then the queue of the best bids eventually empties with probability one, meaning that there is a positive probability that the trading price changes. However, if we
are in the region $\frac{1}{n}\le\tfrac{\lambda_b}{\mu_a n}=\frac{\rho}{n}<1$,
then we know that $A(t),B(t)\to\infty$, 
as $t\to\infty$. This means that the queue of the bids at some price will eventually never empty, which means that trades at lower prices will never occur.
By symmetry, the same argument holds for asks.

\begin{figure}[h]
\begin{center}
\subfloat[]{\includegraphics[width=0.33\columnwidth]{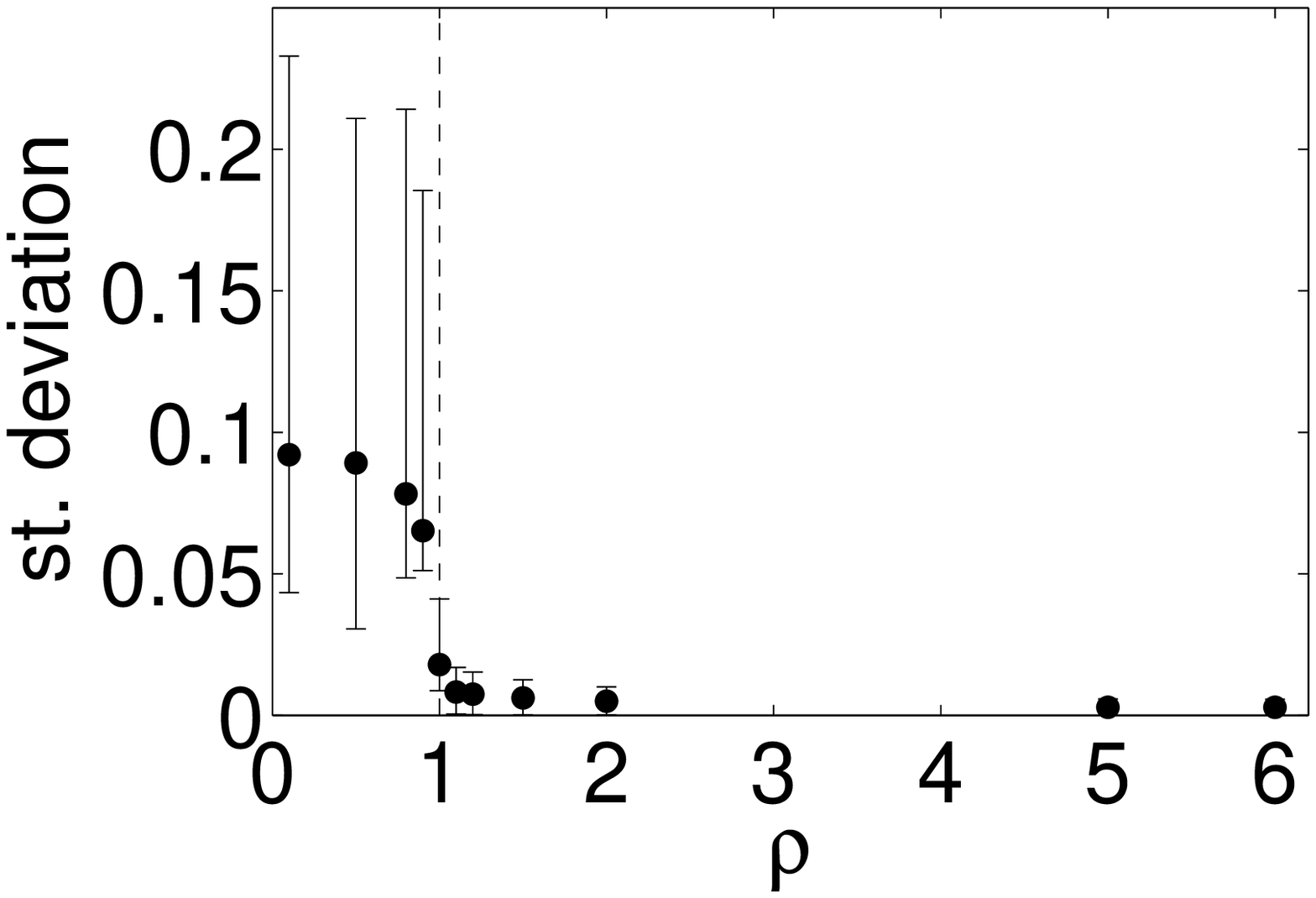}  } 
\subfloat[]{\includegraphics[width=0.33\columnwidth]{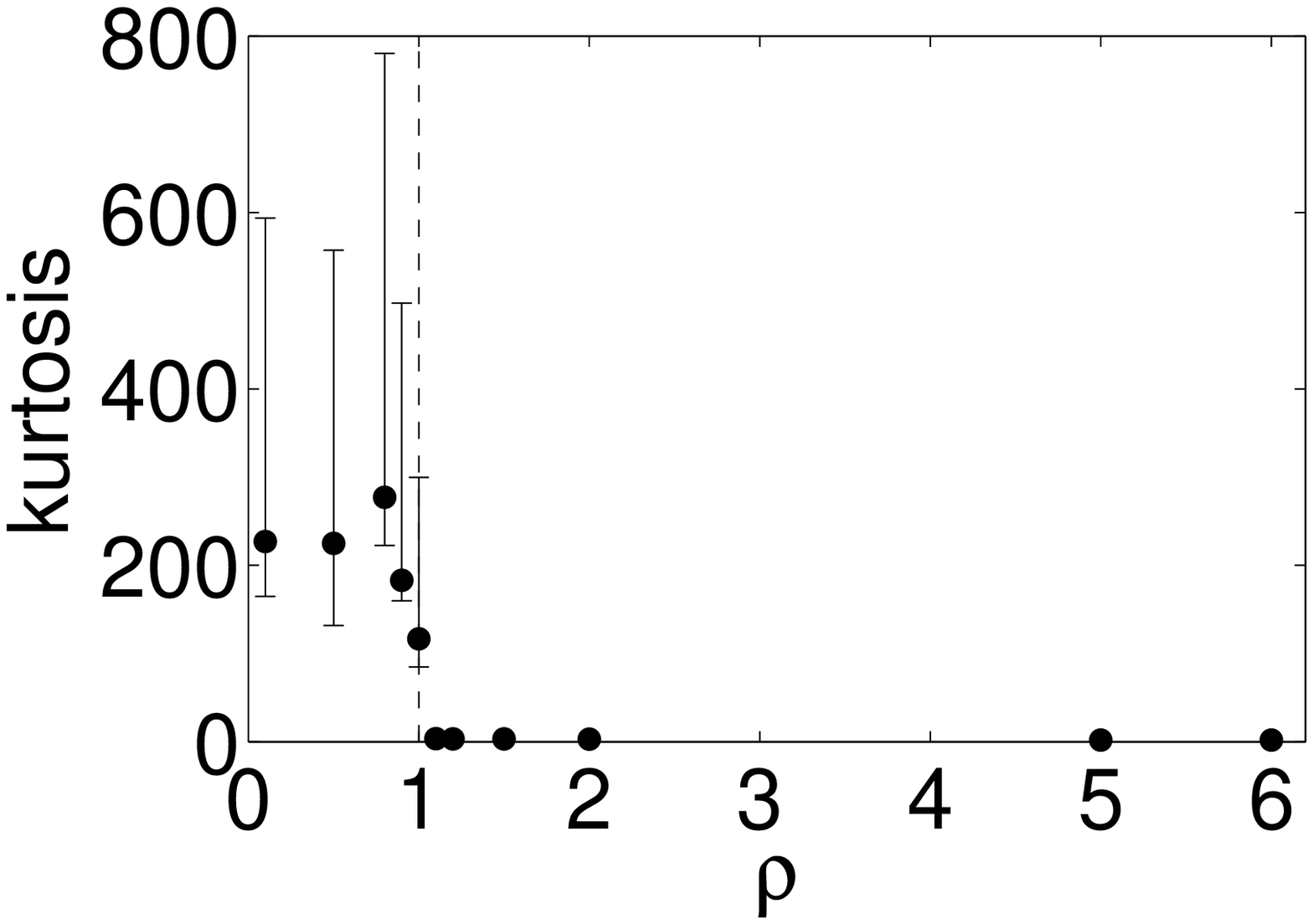}  }
\subfloat[]{\includegraphics[width=0.33\columnwidth]{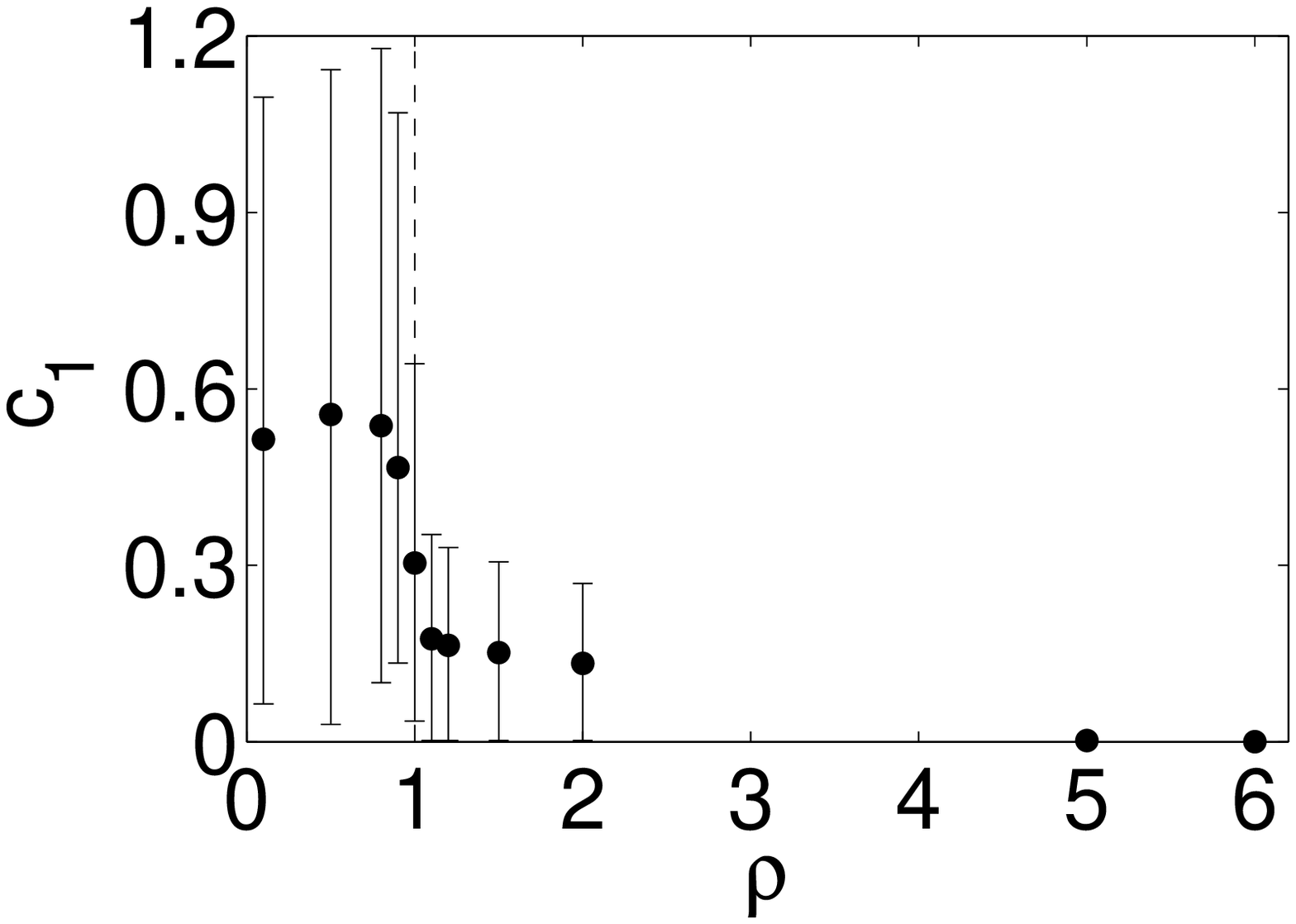}  }
\end{center}
\caption{Mean values and error bars of the standard deviation (a), kurtosis (b) of log-returns and the first-lag autocorrelation of absolute log-returns (c)  as functions of the parameter $\rho$, estimated from 1000 simulations.}
\label{errorbars}
\end{figure}

\begin{figure}[h!]
\begin{center}
\subfloat[]{\includegraphics[width=0.33\columnwidth]{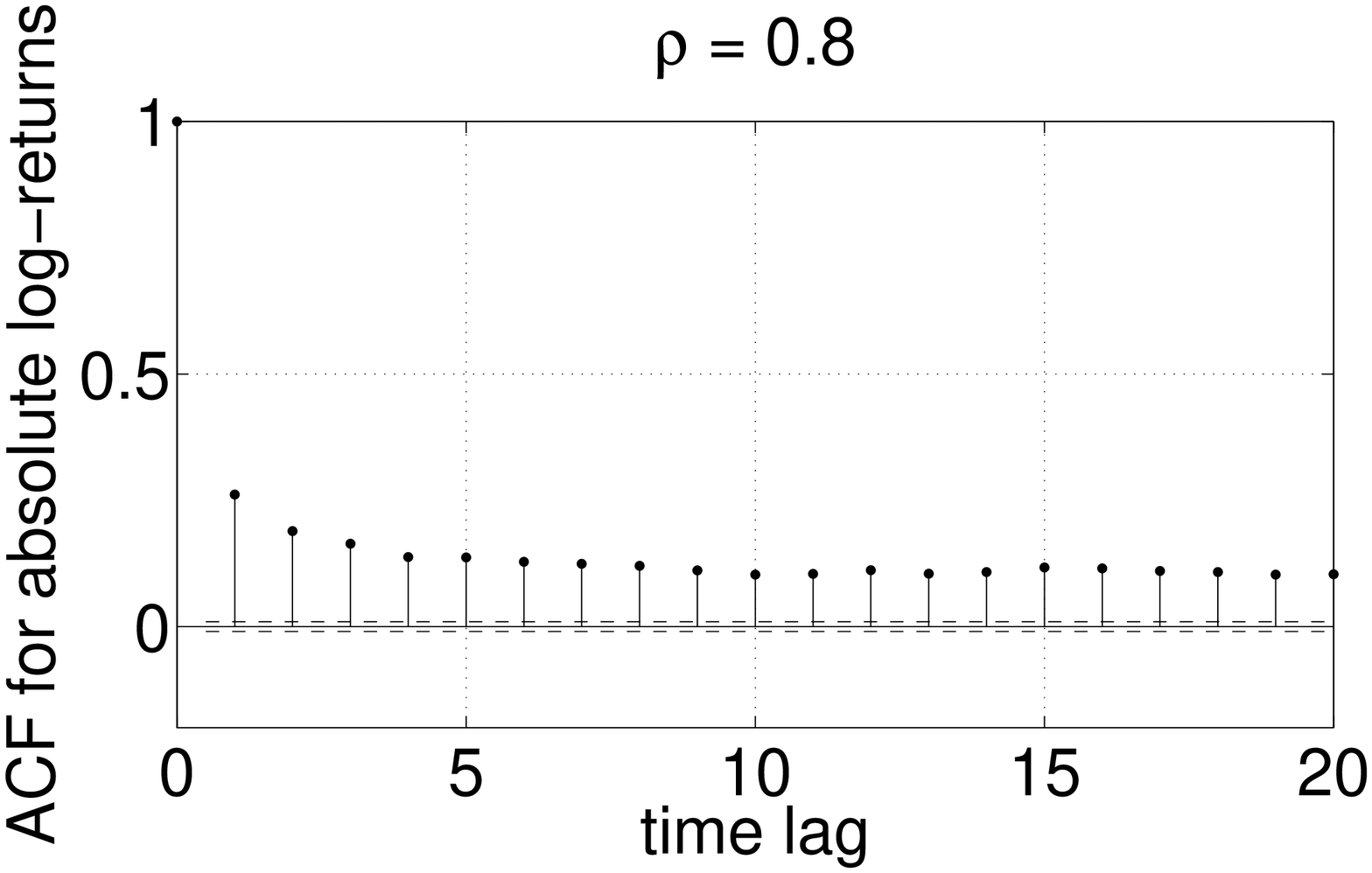} }
\subfloat[]{\includegraphics[width=0.33\columnwidth]{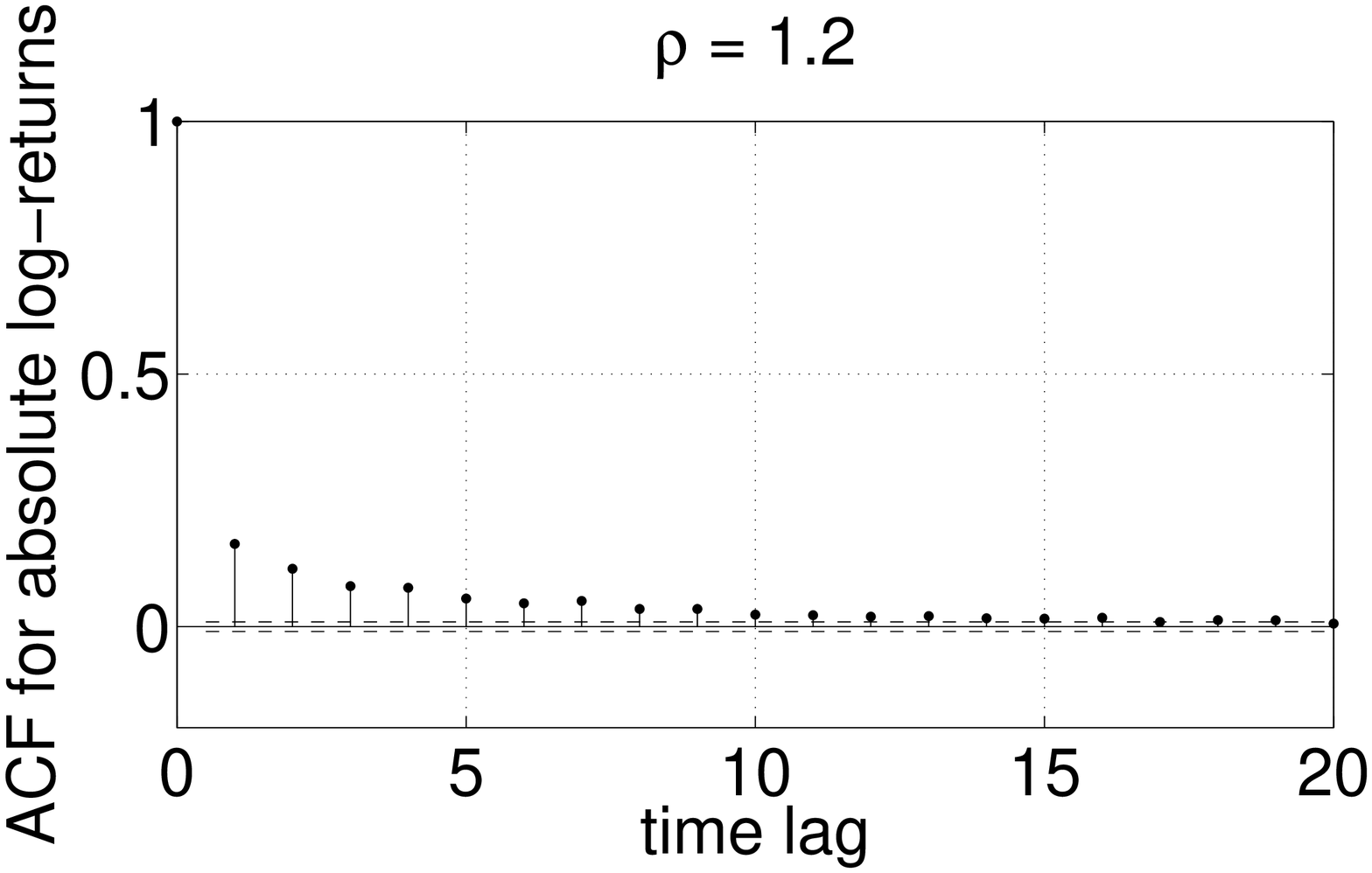} } 
\subfloat[]{\includegraphics[width=0.33\columnwidth]{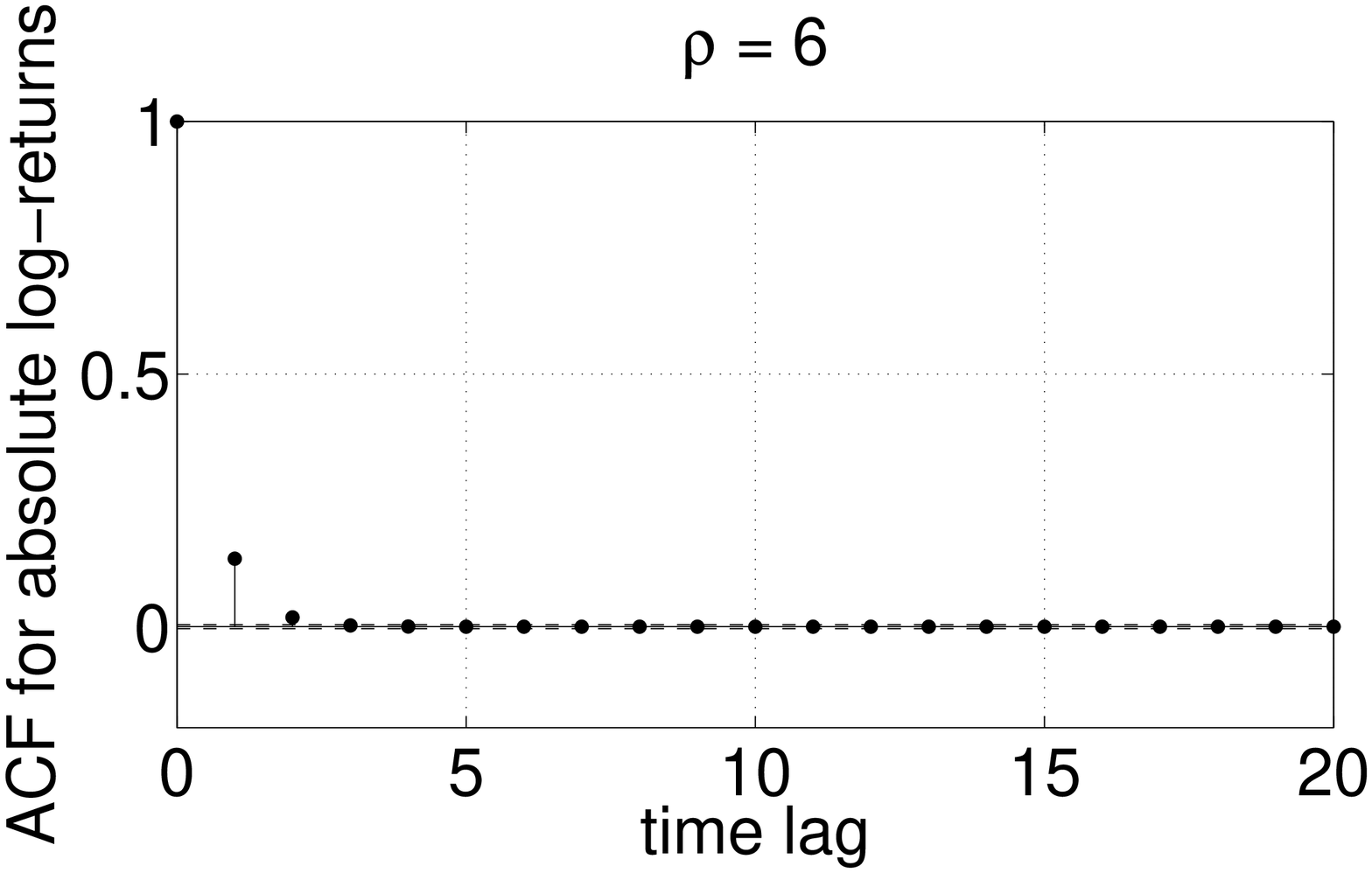} } 
\caption{Sample autocorrelation functions for the absolute log-returns. (a) ergodic case ($\rho=0.8$); (b) non-ergodic case for $1\leq \rho <n$ ($\rho=1.2$) and (c) non-ergodic case for $\rho \geq n$ ($\rho=6$).} 
\label{acf}
\end{center}
\end{figure}

\begin{figure}[h]
\begin{center}
\subfloat[]{\includegraphics[width=0.5\columnwidth]{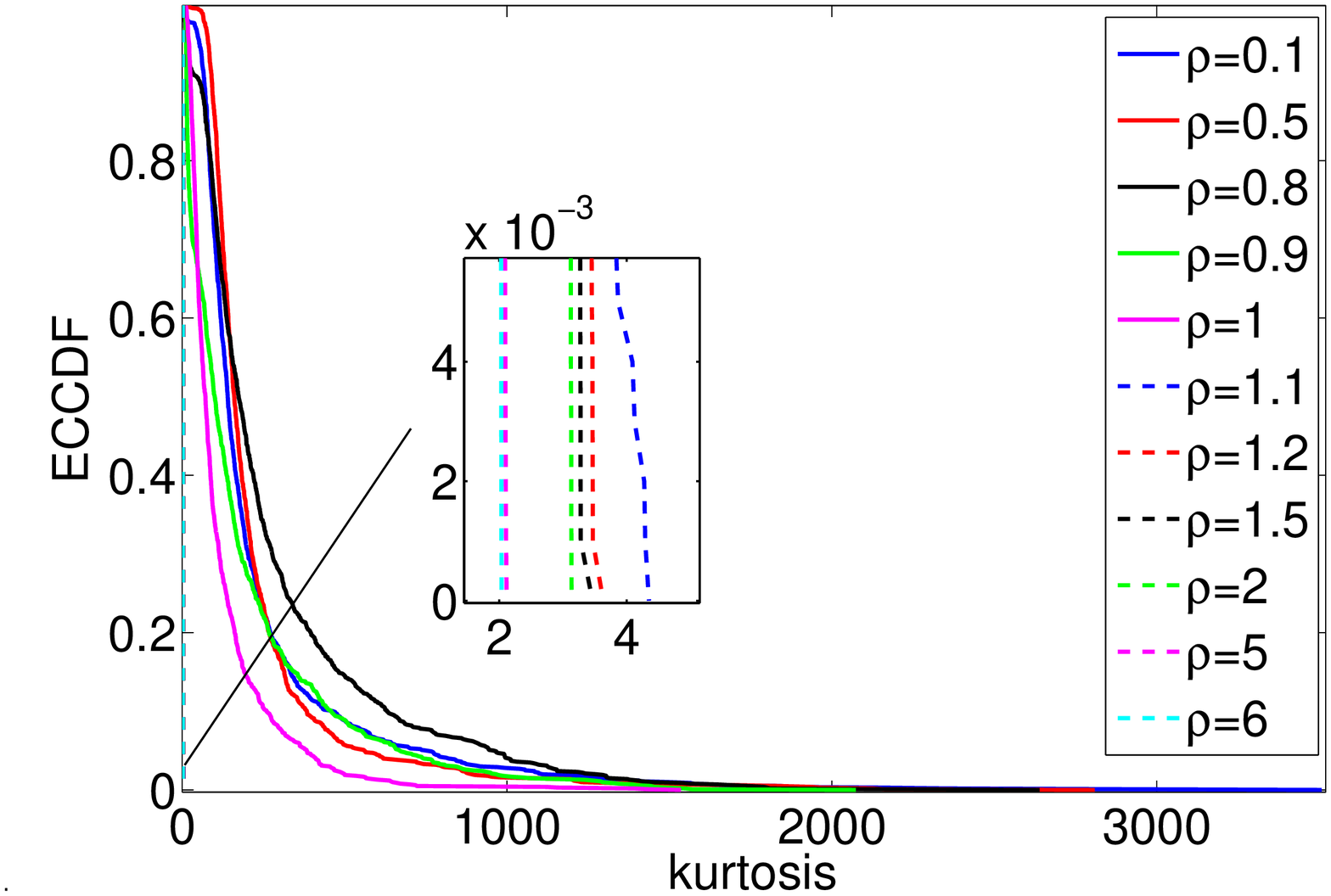} }
\subfloat[]{\includegraphics[width=0.5\columnwidth]{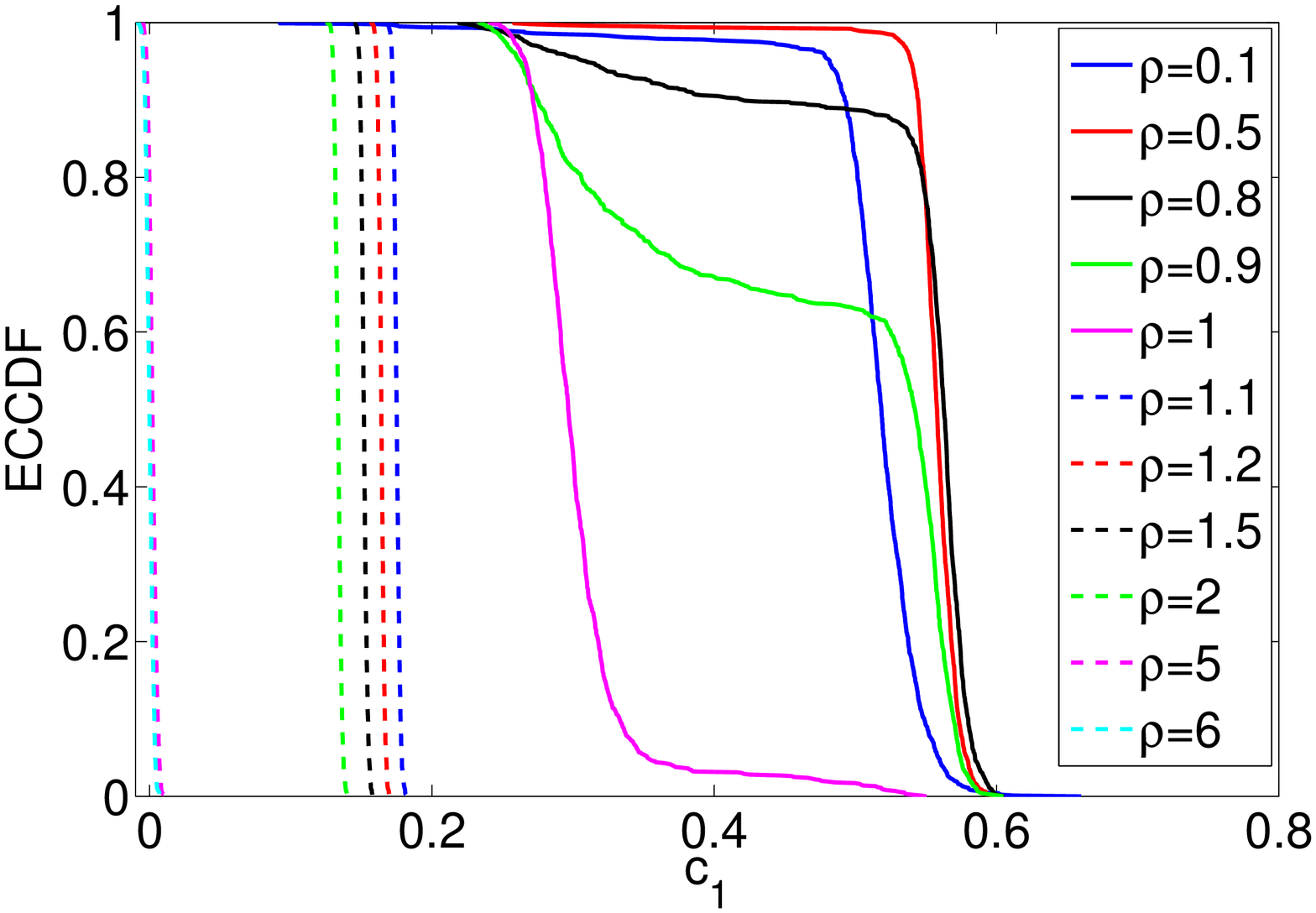} }
\caption{(Color online) Empirical complementary cumulative distribution functions of the kurtosis of logarithmic returns (a) and the autocorrelation at the first-lag of absolute log-returns (b) for different values of $\rho$.}
\label{kurtosis}
\end{center}
\end{figure}

The transition between regimes can be detected studying the moments of log-returns.
In  table \ref{stat}, we give the descriptive statistics for 
log-returns, including mean, standard deviation, skewness and kurtosis as well as the autocorrelation coefficient at the first lag  of absolute log-returns, $c_1$, for different values of the parameter $\rho$. These statistics are computed on 1000 simulation runs with $10^6$ events.
\begin{table}[h]
\begin{center}
\begin{tabular}{|c|ccccc|} 
\hline
 & $\rho=0.1$  &$\rho=0.8$ &$\rho=1$  & $\rho=1.2$ & $\rho=6$ \\ \hline 
mean & $-3.298\cdot 10^ {-6}$&$-7.472\cdot 10^ {-7}$ &$-2.068\cdot 10^ {-7}$ & $-1.918\cdot 10^ {-9}$&$-3.294\cdot 10^ {-9}$ \\ 
st.\ dev. & 0.0921&0.0782 &0.0179 &0.0075 &0.0028  \\
skewness & 0.896&0.287 &-0.034 & $3.443\cdot 10^{-4}$& $-3.8\cdot 10^{-7}$ \\ 
kurtosis &227.089 &277.33 &116.679 &3.373  &2.004 \\ 
$c_1$ &0.515 &0.538 &0.304 &0.164 &$1.06\cdot 10^{-5}$ \\ \hline
\end{tabular}
\end{center}
\caption{Descriptive statistics of log-returns and the autocorrelation coefficient at the first lag of absolute log-returns for different $\rho$, made on 1000 runs of simulation of $10^6$ events.}
\label{stat}
\end{table}
Figure \ref{errorbars} shows standard deviation, kurtosis and $c_1$ in more detail, namely mean values and error bars are given for these three quantities estimated from 1000 runs. One can see that 
these quantities increase in the non-ergodic case.
Our findings can be compared with those presented in a recent study of financial stylized facts \cite{brandouy}, where the authors find that higher rate of limit
orders stabilizes the market by decreasing the standard deviation of returns.
In Fig.\ \ref{acf}, we plot the sample autocorrelation for the tick-by-tick absolute log-return series for $\rho=0.8$, $\rho=1.2$ and for $\rho=6$. A slow decay of the ACF in ergodic case, showing long
range-memory, is in agreement with
the stylized facts found in financial data. One can see that
this decay is much faster if $\rho$ increases. 
In Fig.\ \ref{kurtosis}, we plot the complementary cumulative distribution of kurtosis values (Fig.\ \ref{kurtosis} (a)) and of $c_1$ (Fig.\ \ref{kurtosis} (b)) for 1000 simulations and for different values of $\rho$ in order to corroborate the observation made above: there is a jump in the kurtosis of logarithmic returns as well as a jump in the first-lag autocorrelation of absolute log-returns as $\rho$ moves from values larger than $1$ to values smaller than $1$.

\section*{Conclusions}

\textcolor{black}{In summary, in this paper we have shown that a symmetric continuous double auction model has three regimes depending on the value of the
parameter $\rho$. If $0 < \rho < 1$, we are in the ergodic regime and prices are free to fluctuate over the full available price range. For $\rho \geq 1$, we
have the transition to a non-ergodic regime which stabilizes prices. However, there is a further transition. If $1 \leq \rho < n$, where $n$ is the size of the allowed range for
limit orders, prices can still fluctuate within a limited range. If $\rho \geq n$, then prices will eventually fluctuate between two values. Even if this is a feature of our simplified model, the distinction between the ergodic and non-ergodic regimes is relevant for real equity markets. Our educated guess, which will be the subject of further research, is that they live in the non-ergodic regime, but not too far from the threshold.}

\section*{Acknowledgments}

We wish to thank Mauro Politi for inspiring discussions. This work was supported by the Italian grant 
PRIN 2009, 2009H8WPX5\_002, {\it Finitary and non-finitary probabilistic methods in economics}.

\end{document}